\documentclass[preprintnumbers,amsmath,amssymb,showpacs]{revtex4}

\usepackage{graphicx}
\usepackage{dcolumn}
\usepackage{bm}
\usepackage{booktabs}
\usepackage{CJK}
\usepackage{subfigure}

\begin{document}

\title{On four-photon entanglement from parametric down-conversion process}

\author{Dong Ding$^{1}$}
\author{Yingqiu He$^{2}$}
\email{yingqiuhe@126.com}
\author{Fengli Yan$^3$ }
 \email{flyan@hebtu.edu.cn}
\author{ Ting Gao$^4$ }
 \email{gaoting@hebtu.edu.cn}
\affiliation {
$^1$ Department of Basic Curriculum, North China Institute of Science and Technology, Beijing 101601, China\\
$^2$ Department of Biomedical Engineering, Chengde Medical University, Chengde 067000, China\\
$^3$ College of Physics Science and Information Engineering, Hebei Normal University, Shijiazhuang 050024, China \\
$^4$ College of Mathematics and Information Science, Hebei Normal University, Shijiazhuang 050024, China}
\date{\today}

\date{\today}

\begin{abstract}

We propose two schemes to generate four-photon polarization-entangled states from the second-order emission of the spontaneous parametric down-conversion process.
By using linear optical elements and the coincidence-detection, the four indistinguishable photons emitted from parametric down-conversion source result in the Greenberger-Horne-Zeilinger (GHZ) state or the superposition of two orthogonal GHZ states.
For this superposition state, under particular phase settings we analyze the quantum correlation function and the local hidden variable (LHV) correlation.
As a result, the Bell inequality derived from the LHV correlation is violated with the visibility larger than 0.442.
It means that the present four-photon entangled state  is therefore suitable for testing the LHV theory.

\end{abstract}

\maketitle

\section{Introduction}

As a quantum resource, multiphoton entanglement \cite{Kok2007, Pan2012} plays an important role  in both theoretical studies and  experimental techniques.
One of the attractive aspects of this field is how to generate the desired multiphoton entangled states \cite{GHZExperiment99, Pan2001, QE-LNumberP2004, WAZ2007, ShengDeng2010b, WLK2011, DYG2013, DYG2014, HDYG2015JPB, HDYG2015OE, Pan-10photon-2017}.
Since a spontaneous parametric down-conversion (PDC) \cite{SPDC1970, PDC1990RT, PDC1992BMM, PDC1995} source is capable of creating pairs of entangled photons,
in general, a standard method of generating the multiphoton entanglement is to evolve the pairs emitted from respective source with a set of passive optical elements \cite{Pan2012}.
For this purpose, it is crucial to suppress undesired multipair emissions and enhance the collection efficiency of the entangled photon pairs.

A higher-order emission of the PDC source \cite{ORW1999, DD2001, LHB2001, SB2003, RSMATZG2004, NOOST2007Science, HDYG2017SD} is related to a twin-beam multiphoton entangled state.
For the second-order emission, it has been shown that \cite{RSMATZG2004}, when the duration of the pump pulse is much shorter than the coherence time of the photons, the emitted state can be described as an indistinguishable twin-beam four-photon entangled state (a genuine four-partite entanglement) rather than two independent pairs.
Based on such four-photon emission, Weinfurter and \.{Z}ukowski  \cite{WZ2001} proposed a novel method of generating
four-photon polarization entangled state.
By using several linear optical elements, a kind of four-photon polarization entangled state can be obtained directly in PDC process, instead of recombining two entangled-photon pairs. This four-photon entangled state can be used to test the local hidden variable (LHV) theory \cite{WZ2001}.
Later, it leads to the following discussion on four-photon entanglement in two-crystal geometry by Li and Kobayashi \cite{LK2004-2005}.

In this paper, we at first focus on the generation of two particular types of four-photon entangled states from the second-order emission of the PDC source.
The output states are quite different from the previous ones because of the interference at an additional polarizing beam splitter (PBS) before the fourfold coincidence detection.
We then compare LHV correlation function and quantum correlation function under particular phase settings.
The result is that the quantum violation of the present four-photon entangled state is much larger than the threshold value 1.

\section{Generation of four-photon polarization entangled states}

\begin{figure}
  \includegraphics[width=2in]{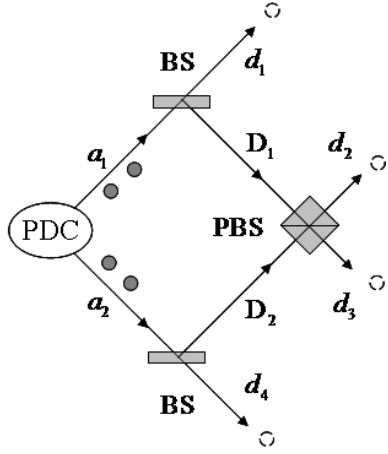}\\
  \caption{The schematic diagram of generating four-photon Greenberger-Horne-Zeilinger state. The parametric down-conversion (PDC) source is used to produce four indistinguishable twin-beam entangled photons. Two $50:50$ beam splitters (BSs) and the following polarizing beam splitter (PBS) are used to evolve these photons from the spatial modes $a_{1}$ and $a_{2}$ to $d_{1}$, $d_{2}$, $d_{3}$ and $d_{4}$.}
  \label{}
\end{figure}

For the second-order emission of the PDC source,
suppose that the four-photon entangled state emitted from the PDC source is
\begin{eqnarray}\label{4-photon-pdc}
|\Phi_\text{}\rangle &=& \frac{1}{2\sqrt{3}}({\hat a_{1_H}^{\dag}}{\hat a_{2_V}^{\dag}} - {\hat a_{1_V}^{\dag}}{\hat a_{2_H}^{\dag}})^{2}|0\rangle,
\end{eqnarray}
where $\hat a^{\dag}_{i_{H}}$ and $\hat a^{\dag}_{i_{V}}$ (with $i=1,2$) are respectively the creation operators with horizontal and vertical polarization in the spatial modes $a_i$.
We next describe a method of generating two four-photon polarization entangled states, i.e., the Greenberger-Horne-Zeilinger (GHZ) state and the superposition of two orthogonal GHZ states.

At first, we show the scheme of generating the four-photon GHZ state.
As indicated  in Fig.1, we here assume that the two polarization independent $50:50$ beam splitters (BSs) transform $a_{1}$ into $(d_{1}+D_{1})/\sqrt{2}$ and transform $a_{2}$ into $(d_{4}-D_{2})/\sqrt{2}$, respectively.
Since the interference at the two BSs, the initial twin-beam four-photon entangled state evolves
\begin{eqnarray}\label{}
|\Phi_\text{BS}\rangle &=&
\frac{1}{4\sqrt{3}}
  [(|2,0\rangle_{d_{1_H}D_{1_H}} + |0,2\rangle_{d_{1_H}D_{1_H}} + \sqrt{2}|1,1\rangle_{d_{1_H}D_{1_H}})\otimes
   (|2,0\rangle_{d_{4_V}D_{2_V}} + |0,2\rangle_{d_{4_V}D_{2_V}} - \sqrt{2}|1,1\rangle_{d_{4_V}D_{2_V}})
\nonumber \\ & &
+  (|2,0\rangle_{d_{1_V}D_{1_V}} + |0,2\rangle_{d_{1_V}D_{1_V}} + \sqrt{2}|1,1\rangle_{d_{1_V}D_{1_V}})\otimes
   (|2,0\rangle_{d_{4_H}D_{2_H}} + |0,2\rangle_{d_{4_H}D_{2_H}} - \sqrt{2}|1,1\rangle_{d_{4_H}D_{2_H}})
\nonumber \\ & &
- |1,1\rangle_{(d_{1_H}d_{1_V} + d_{1_H}D_{1_V} + d_{1_V}D_{1_H} + D_{1_H}D_{1_V})} \otimes
  |1,1\rangle_{(d_{4_H}d_{4_V} - d_{4_H}D_{2_V} - d_{4_V}D_{2_H} + D_{2_H}D_{2_V})}].
\end{eqnarray}
Here, for simplicity, $|m,n\rangle_{d_{1_H}D_{1_V}}$, for example, represents that there are $m$ horizontally polarized photons in spatial mode $d_{1}$ and $n$ vertically polarized photons in spatial mode $D_{1}$.
When photons travel  through the PBS in the horizontal-vertical basis,
the evolution of the photons from the spatial modes $D_{1}$ and $D_{2}$ to $d_{2}$ and $d_{3}$ is given by
\begin{equation}\label{transformation-D2d}
|V\rangle_{D_{1}} \rightarrow |V\rangle_{d_{2}},~~~  |H\rangle_{D_{1}} \rightarrow |H\rangle_{d_{3}},~~~
|V\rangle_{D_{2}} \rightarrow |V\rangle_{d_{3}},~~~  |H\rangle_{D_{2}} \rightarrow |H\rangle_{d_{2}}.
\end{equation}
At last, if we only  consider a four-photon coincidence detection, i.e., there is one photon in each of the spatial modes $d_{1}, d_{2}, d_{3}$ and $d_{4}$, then we have
\begin{equation}\label{}
 |\psi\rangle=\frac{1}{\sqrt{2}}(|HVVH\rangle_{d_{1}d_{2}d_{3}d_{4}} + |VHHV\rangle_{d_{1}d_{2}d_{3}d_{4}}).
\end{equation}
Obviously, this output state is exactly the canonical four-photon polarization-entangled GHZ state.

\begin{figure}
  \includegraphics[width=2.2in]{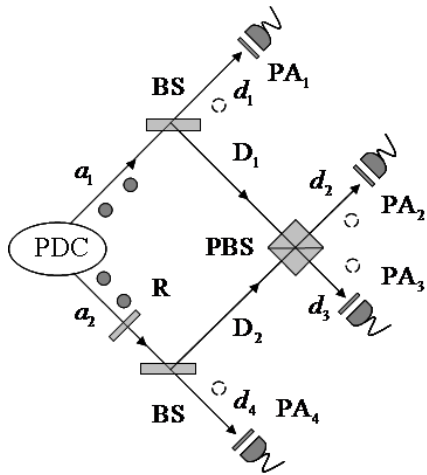}\\
  \caption{Generation and detection of the four-photon polarization entangled state. $\text{R}_{}$ represents a half wave plate, which is used to convert the polarization state $|H\rangle$ into $|V\rangle$ or vice versa. Each $\text{PA}_{i}~(i=1,2,3,4)$ is a polarization analysis used to investigate the entanglement property of the output four-photon entangled state.}
  \label{}
\end{figure}

We next extend the above method to exploring an available four-photon polarization entangled state.
For this purpose, let us first insert a half wave plate $\text{R}_{}$ in spatial mode $a_{2}$, which can convert the polarization state $|H\rangle$ into $|V\rangle$ or vice versa.
Fig. 2 shows the schematic diagram of generating this kind of four-photon polarization entangled state.
  Due to the action of this half wave plate, the original four-photon entangled state (\ref{4-photon-pdc}) becomes
\begin{eqnarray}\label{}
|\Phi_\text{R}\rangle = \frac{1}{2\sqrt{3}}({\hat a_{1_H}^{\dag}}{\hat a_{2_H}^{\dag}} - {\hat a_{1_V}^{\dag}}{\hat a_{2_V}^{\dag}})^{2}|0\rangle.
\end{eqnarray}

Similarly, when the photons pass through two BSs, one has
\begin{eqnarray}\label{}
|\Phi_\text{BS}\rangle &=&
\frac{1}{4\sqrt{3}}
  [(|2,0\rangle_{d_{1_H}D_{1_H}} + |0,2\rangle_{d_{1_H}D_{1_H}} + \sqrt{2}|1,1\rangle_{d_{1_H}D_{1_H}})\otimes
   (|2,0\rangle_{d_{4_H}D_{2_H}} + |0,2\rangle_{d_{4_H}D_{2_H}} - \sqrt{2}|1,1\rangle_{d_{4_H}D_{2_H}})
\nonumber \\ & &
+  (|2,0\rangle_{d_{1_V}D_{1_V}} + |0,2\rangle_{d_{1_V}D_{1_V}} + \sqrt{2}|1,1\rangle_{d_{1_V}D_{1_V}})\otimes
   (|2,0\rangle_{d_{4_V}D_{2_V}} + |0,2\rangle_{d_{4_V}D_{2_V}} - \sqrt{2}|1,1\rangle_{d_{4_V}D_{2_V}})
\nonumber \\ & &
- |1,1\rangle_{(d_{1_H}d_{1_V} + d_{1_H}D_{1_V} + D_{1_H}d_{1_V} + D_{1_H}D_{1_V})} \otimes
  |1,1\rangle_{(d_{4_H}d_{4_V} - d_{4_H}D_{2_V} - D_{2_H}d_{4_V} + D_{2_H}D_{2_V})}].
\end{eqnarray}
Then, applying the transformations (\ref{transformation-D2d}) and conditioning on detecting one photon in each of the four spatial modes,
 we have the four-photon polarization entangled state
\begin{equation}\label{psi-GHZ-2-1}
 |\psi\rangle=\frac{1}{\sqrt{10}}[(|HVVH\rangle_{d_{1}d_{2}d_{3}d_{4}} + |VHHV\rangle_{d_{1}d_{2}d_{3}d_{4}}) - 2(|HHHH\rangle_{d_{1}d_{2}d_{3}d_{4}} + |VVVV\rangle_{d_{1}d_{2}d_{3}d_{4}})].
\end{equation}
Obviously, instead of the GHZ state, it is the superposition of two orthogonal GHZ states with unequal probabilities.

\section{Quantum nonlocality of the four-photon entangled state}

As is well known, Bell inequalities can be used to investigate the constitutive relations for multiparticle entanglement \cite{HHHH2009, GT2009, Peres1996PRL77-1413, DVC2000, YGC2011, GYE2014} and quantum nonlocality \cite{Bell1964, CHSH1969, Mermin1990, Ardehali1992, BK1993, WW2001, ZB2002, HDYG2015EPL}.
We now turn to the discussion of quantum nonlocality of the present four-photon entangled state.
Four-photon GHZ state maximally violates Mermin inequality \cite{Mermin1990}.  Especially for a four-particle system,
it has been shown that \cite{HDYG2015EPL} a compact Mermin-type inequality is also maximally violated by four-qubit GHZ state with a certain constant visibility 2.
Next we  concentrate on quantum nonlocality of the aforementioned superposition state (\ref{psi-GHZ-2-1}).

For a four-photon system, for the sake of simplicity,  we suppose that each location of the photon is allowed to choose independently between two dichotomic observables and each outcome can take one of the values $+1$ or $-1$.
In order to investigate quantum nonlocality of the present four-photon entangled state, we here introduce a polarization analysis basis \cite{WZ2001}
\begin{equation}\label{}
|m_{x},\phi_{x}\rangle = \frac{1}{\sqrt{2}}(|V\rangle_{x} + m_{x} \text{e}^{-\text{i}\phi_{x}} |H\rangle_{x}),
\end{equation}
where $\phi_{x}$ is a local phase setting connecting with location $d_{x}$ ($x=1,2,3,4$), and $m_{x}=\pm1$ are two possible measurement results.

The correlation function of measurement results is usually defined as the average of the product of the local values.
Then, the correlation function for the present four-photon entangled state (\ref{psi-GHZ-2-1}) is
\begin{eqnarray}\label{E-cc}
E (\phi_1,\phi_2,\phi_3,\phi_4)&=& \sum_{m_1,m_2,m_3,m_4 = \pm 1} m_1m_2m_3m_4 |\langle m_1,\phi_1|\langle m_2,\phi_2|\langle m_3,\phi_3|\langle m_4,\phi_4|\psi\rangle|^{2} \nonumber \\
 &=& \frac{4}{5} \text{cos}(\phi_1+\phi_2+\phi_3+\phi_4) + \frac{1}{5} \text{cos}(\phi_1-\phi_2-\phi_3+\phi_4).
\end{eqnarray}

On the other hand, in terms of the local hidden variable (LHV) model, the photons are locally but realistically correlated. A four-photon correlation function for two alternative dichotomic measurements can be described as a four-index tensor \cite{WZ2001}
\begin{equation}\label{}
\hat{E}_{\text{LHV}}
=\sum_{k_1,k_2,k_3,k_4 = 1,2} c_{k_1,k_2,k_3,k_4} \textbf{v}_{1}^{k_1} \otimes \textbf{v}_{2}^{k_2} \otimes \textbf{v}_{3}^{k_3}\otimes \textbf{v}_{4}^{k_4},
\end{equation}
where $\textbf{v}_{x}^{k_{x}}$ ($x=1,2,3,4$) represents a two-dimensional real vector with $k_x = 1,2$ and  $\textbf{v}_{x}^{1}=(1,1),\textbf{v}_{x}^{2}=(1,-1)$,
and $c_{k_1,k_2,k_3,k_4}$ are  the correlation coefficients satisfying
\begin{eqnarray}\label{BI}
\sum_{k_1,k_2,k_3,k_4 = 1,2} |c_{k_1,k_2,k_3,k_4}| \leq 1.
\end{eqnarray}
This inequality is derived from LHV model and generically called Bell inequality, which can be used to test the LHV theory.

In order to compare the LHV correlation and the quantum correlation for the present state (\ref{psi-GHZ-2-1}), we now rewrite the quantum correlation function as a tensor
\begin{equation}\label{}
\hat{E}_{\text{Q}}
=\sum_{k_1,k_2,k_3,k_4 = 1,2} q_{k_1,k_2,k_3,k_4} \textbf{v}_{1}^{k_1} \otimes \textbf{v}_{2}^{k_2} \otimes \textbf{v}_{3}^{k_3}\otimes \textbf{v}_{4}^{k_4}.
\end{equation}
For simplicity, we here take the experimental settings
$\phi_{1}^{1}=0$, $\phi_{1}^{2}=\pi/2$, $\phi_{2,3,4}^{1}= - \pi/4$ and $\phi_{2,3,4}^{2}=\pi/4$, as described in \cite{WZ2001}.
The quantum correlation coefficients $q_{k_1,k_2,k_3,k_4}$ can be obtained straightforwardly by using equation (\ref{E-cc}), and then we have
\begin{eqnarray}\label{}
\sum_{k_1,k_2,k_3,k_4 = 1,2} |q_{k_1,k_2,k_3,k_4}| = \frac{16}{5\sqrt{2}}.
\end{eqnarray}
This result shows that the Bell inequality (\ref{BI}) derived from the LHV correlation is violated with the value $8\sqrt{2}/5$, which is much larger than the threshold value 1.

\section{discussion and summary}

In summary, we have shown two schemes of generating four-photon polarization entangled states and discussed quantum nonlocality of the present four-photon entangled state.
For a four-photon emission of the PDC source,
due to the interference at PBS before the fourfold coincidence detection, we can obtain the GHZ state or the superposition of two orthogonal GHZ states.
Since only a few linear optical elements and the coincidence-detection measurements are employed, it is convenient to realize in experiments.

For quantum nonlocality of the present superposition of two orthogonal GHZ states,
by calculating the quantum correlation function under particular phase settings, it has been shown that the present four-photon entangled state violates the Bell inequality with the visibility $5\sqrt{2}/16 \simeq 0.442$, which is superior to the visibility $3\sqrt{2}/8 \simeq 0.53$ associated with the superposition of the GHZ state and the product state of two Bell states \cite{WZ2001}.
This means that the present four-photon entangled state is well suited to testing the quantum formalism against the LHV model in experiments.

{\it Acknowledgements. }
This work was supported by the National Natural Science Foundation of China under Grant Nos: 11475054, 11371005, 11547169, Hebei Natural Science Foundation of China under Grant No: A2016205145, the Research Project of Science and Technology in Higher Education of Hebei Province of China under Grant No: Z2015188, Foundation for High-level Talents of Chengde Medical University under Grant No: 201701, Langfang Key Technology Research and Development Program of China under Grant No: 2014011002.

\end{document}